\documentclass{article}

\usepackage{arxiv}

\usepackage[utf8]{inputenc} % allow utf-8 input
\usepackage[T1]{fontenc}    % use 8-bit T1 fonts
\usepackage{hyperref}       % hyperlinks
\usepackage{url}            % simple URL typesetting
\usepackage{booktabs}       % professional-quality tables
\usepackage{amsfonts}       % blackboard math symbols
\usepackage{nicefrac}       % compact symbols for 1/2, etc.
\usepackage{microtype}      % microtypography
\usepackage{lipsum}		% Can be removed after putting your text content
\usepackage{graphicx}
\usepackage{doi}
\usepackage{bm}
\usepackage{amssymb}
\usepackage{float}
\usepackage{amsmath}

\usepackage{physics}
\usepackage[table,xcdraw]{xcolor}

\title{Influence of higher order electron-phonon interaction on the electron-related lattice thermal properties of 2D Dirac crystals}

\date{November 1, 2023}	% Here you can change the date presented in the paper title
\author{ \href{}{\hspace{1mm} Sina Kazemian\textsuperscript{1}, Giovanni Fanchini\textsuperscript{2}}\\
 \textsuperscript{1}Department of Physics and Astronomy, Western and Waterloo University, \texttt{s4kazemi@uwaterloo.ca }\\
 \textsuperscript{2}Department of Physics and Astronomy, Western Ontario University, \texttt{gfanchin@uwo.ca}
}
\renewcommand{\shorttitle}{$2^{nd}$ order e-ph interactions impact on electron-related thermal properties of 2D Dirac crystals}

\begin{document}
\maketitle

\begin{abstract}

Understanding the numerous crucial properties of Dirac crystals, such as their thermal conductivity, necessitates the use of models that consider the interaction between Dirac electrons and persistent acoustic phonons in which the oscillation frequency $\omega$ depends on the phonon wave vector $q$ and is therefore dispersive. It is commonly assumed that the exceptionally high thermal conductivity of two-dimensional (2D) Dirac crystals is due to the near ideality of their phonon quantum gasses with undesired limitations originating from phenomena such as electron-phonon (e-ph) interactions. Electrons transferred to Dirac crystals from metal nano-particles through doping have been shown to affect and limit the thermal conductivity of Dirac crystals due to e-ph interactions at distances up to several microns from the nano-particle. Notably, the e-ph thermal conductivity is directly linked to the phonon scattering rate, demonstrating a proportional relationship. Customarily, when calculating the phonon scattering rate, it is common to overlook phonons with short-dispersive wavelengths since in metals $q$ is significantly smaller than the Fermi surface dimensions. However, this approach proves insufficient for analyzing 2D Dirac crystals. Furthermore, the in-plane phonon scattering rate is calculated up to the first order of magnitude consisting of 2 electrons and 1 phonon, 3-particle interaction. In these calculations, only processes involving the decay of an electron and phonon, leading to the creation of a new electron (EP-E*) are considered. However, processes that involve the decay of an electron and the creation of a new electron and phonon (E-E*P*) are not taken into consideration. In this paper, we present an accurate expression for the phonon scattering rate and the e-ph thermal conductivity in 2D Dirac crystals for in-plane phonons considering phonons with short-dispersive wavelengths. We further demonstrate that even at room temperatures, when calculating the phonon scattering rate and e-ph thermal conductivity, in the case of first-order e-ph interactions, the E-E*P* process assumes significance. In the end, we show the importance of incorporating second-order e-ph interactions, particularly the (EP-E*P*) interaction involving the decay of an electron and phonon and the creation of a new pair for in-plane phonons, when determining the phonon scattering rate and e-ph thermal conductivity at high temperatures and low Fermi energies. This 4-particle interaction process proves significant in accurately characterizing these properties.

%2-electron, 1-phonon, three particle-interactions vs. 2-electron, 2-phonon. four particle-interactions and their impact on the thermal conductivity of 2D Dirac crystals

\end{abstract}

%\keywords{Suggested keywords}%Use showkeys class option if keyword
                              %display desired
\maketitle

%\begin{quotation}
%The ``lead paragraph'' is encapsulated with the \LaTeX\ 
%\verb+quotation+ environment and is formatted as a single paragraph before %the first section heading. 
%(The \verb+quotation+ environment reverts to its usual meaning after the %first sectioning command.) 
%Note that numbered references are allowed in the lead paragraph.
%%
%The lead paragraph will only be found in an article being prepared for the %journal \textit{Chaos}.
%\end{quotation}

\section{\label{sec:level1}Introduction}

Dirac crystals are a class of zero band-gap solids distinguished by their unique electronic structure: a linear band configuration within the crystal momentum. This stands in stark contrast to the prevalent quadratic band structure typically encountered near the Fermi level in metals and semiconductors \cite{wehling2014dirac,wang2015rare,cayssol2013introduction,fu20192d}. As a result of this, dispersionless electrons exhibit a behavior similar to photons on the Dirac light cone in relativity \cite{dirac1928quantum}. The high charge mobility of Dirac crystals can exhibit exotic quantum phenomena such as the quantum Hall effect, enriching their unique properties and scientific appeal \cite{tajima2013quantum,yuan2016quantum,calizo2007variable}. Furthermore, undoped Dirac crystals have their valence and conduction bands meet at the K-point of the Brillouin-zone boundary as well as the Fermi energy $E_{\mathrm{F}}$ and as a result, the Fermi surface of the crystal degenerates into a single point of the electronic band-structure \cite{wehling2014dirac,wang2015rare}. In the case of two-dimensional (2D) systems, this degeneracy allows for the tuning of the electronic state density at $E_{\mathrm{F}}$, as well as the system's thermal conductivity, through the use of external electric fields or tunable doping \cite{yan2007electric}. As a result, even slight fluctuations of $E_{\mathrm{F}}$ can cause a significant increase in the carrier density by several orders of magnitude \cite{novoselov2007electronic,li2018review}. Such Fermi-level shifts are also anticipated to have a substantial impact on the interaction strength between charge carriers and lattice phonons. This tunable aspect of Dirac crystals has profound implications, particularly in the realm of the electron-phonon (e-ph) interaction \cite{roy2014migdal,hu2021phonon}. It extends to the field of thermal conductivity  \cite{kazemian2017modelling,balandin2020phononics}, where the ability to manipulate $E_{\mathrm{F}}$ can potentially result in tailored thermal properties, offering exciting opportunities for both fundamental research and practical applications \cite{malekpour2016thermal,khokhriakov2018tailoring,politano2018tailoring}.

For nearly a century the understanding of the thermal and electronic properties of condensed matter systems relied on the separation of contributions from electrons and phonons \cite{born1927quantentheorie}, with e-ph interaction treated only as a small perturbation \cite{lee2020ab,marini2015many}. Likewise, this approach, based on the adiabatic Born-Oppenheimer approximation \cite{born1927quantentheorie} has been a cornerstone for understanding the thermal conductivity of solids. Within this approach, there are two pathways for non-zero thermal transport throughout the solid. One is by phonons which are ion-core vibrations in a crystal lattice mostly present in insulators and the second, is by electrons mostly present in metals and other electrical conducting materials. Nevertheless, this approach overlooks the intertwined contributions of electrons and phonons in heat transport. The crucial role of the Born-Oppenheimer approximation is evident in various methodologies, including the application of ab initio solvers for the electron and phonon Boltzmann transport equations \cite{protik2022elphbolt, li2015electrical}. Notably, the breakdown of the Born-Oppenheimer approximation has been recently affirmed in a novel class of 2D quantum solids \cite{pisana2007breakdown}, with graphene, a Dirac crystal, serving as a progenitor for these medium-correlated quantum systems \cite{keimer2017physics, calderon2020correlated}. Consequently, a more profound comprehension of e-ph interactions in Dirac crystals becomes imperative. While the big picture of e-ph interaction in solids is clear enough the details in 2D Dirac crystals are more complicated and difficult to grasp. This includes an accurate calculation of the screening potential and considering phonon creation processes as well as going to higher order e-ph interactions. The first-order e-ph interaction for in-plane phonons is a 3-particle process consisting of two electrons and one phonon, while the second-order e-ph interaction is a 4-particle process consisting of two electrons and two phonons.

The significance of e-ph interaction in lattice conduction dates back to the 1930s and underwent further examination in 1956 by \textit{Ziman et al.} \cite{ziman1957effect}. Subsequent to this seminal work, both experimental and theoretical studies continued to contribute to our understanding of this critical phenomenon \cite{butler1978electron,liao2015significant,kim2016electronic,yang2021indirect,borysenko2010first}. By calculating the phonon scattering rate \textit{Ziman} derived an expression for the e-ph thermal conductivity and showed that the thermal conductivity of a solid decreases due to a 3-particle e-ph interaction process where an electron and phonon decay and a new electron gets created (EP-E*). In this approach, the computation carried out to derive the phonon scattering rate commonly assumes acoustic phonons with long wavelengths \cite{butler1978electron,liao2015significant}. This particular assumption is tailored for metals possessing extensive Fermi surfaces near $E_{\mathrm{F}}$, resulting in $q$ values significantly smaller than the Fermi surface dimensions which imply a short screening length \cite{maldague1978many}. On the contrary, in undoped 2D Dirac crystals, the Fermi surface area collapses to zero resulting in the screening length tending to infinity. Therefore, in order to have an accurate expression for the e-ph thermal conductivity of 2D Dirac crystals the phonon scattering rate should be written for phonons with short-dispersive energy wavelengths. Furthermore, this approach neglects the 3-particle e-ph interaction process where an electron decays and a new electron and phonon get created (E-E*P*). This approximation is correct in metals where we have a large free electron density in the proximity of the Fermi energy $E_\mathrm{F}$ in which $E_\mathrm{F}>>k_B \mathrm{T}$, leading to the transition rate of the EP-E* process to be much larger than the E-E*P* process. However, this is not true in 2D Dirac crystals which have a limited concentration of free electrons and we cannot generally assume $E_\mathrm{F}$ to be much larger than $k_B \mathrm{T}$. Under these circumstances, it is crucial to take into account the E-E*P* process, as it influences the thermal conductivity of the 2D Dirac crystal. This is due to the fact that the phonons created in the E-E*P* process partially or completely offset the phonons eliminated in the EP-E* process, resulting in an alteration of the thermal conductivity. This further leads the study of the 4-particle e-ph interaction necessary in which an electron and a phonon decay and a new pair of electron and phonon get created (EP-E*P*).

This paper introduces a theoretical framework aimed at computing the electron-related lattice thermal conductivity of 2D Dirac crystals, with a particular emphasis on in-plane e-ph interactions and the implications of the Umklapp process on these interactions. The analysis focuses on in-plane phonons and incorporates the intricacies of the 3-particle and 4-particle electron-acoustical phonon interaction processes. To this end, we derive an accurate expression of the phonon scattering rate for phonons with short-dispersive energy wavelengths. We then proceed to write the phonon scattering rate and e-ph thermal conductivity for the 3-particle and 4-particle e-ph interaction processes and compare them with each other. We delve into the open problem of examining the influence of Umklapp scattering\cite{ashcroft2022solid} on the e-ph thermal conductivity of 2D Dirac crystals at elevated temperatures and the intricate connection between e-ph thermal conduction and phonon lifetimes across diverse 2D Dirac materials, shedding light on an uncharted facet of their thermal properties. Our findings reveal a notable shift in dominance within the e-ph interaction processes. Specifically, at elevated Fermi energies and lower temperatures, the 3-particle process takes precedence. Conversely, with an increase in temperature (T>300 K) and a reduction in Fermi energy, the significance of the 4-particle e-ph interaction becomes more pronounced. This transition bears considerable significance in the accurate computation of phonon scattering rates and electron-phonon thermal conductivity in 2D Dirac crystals.

\section{\label{sec:level2}Methodology}

\begin{figure}[h]
\centering
\includegraphics[width=.6\columnwidth]{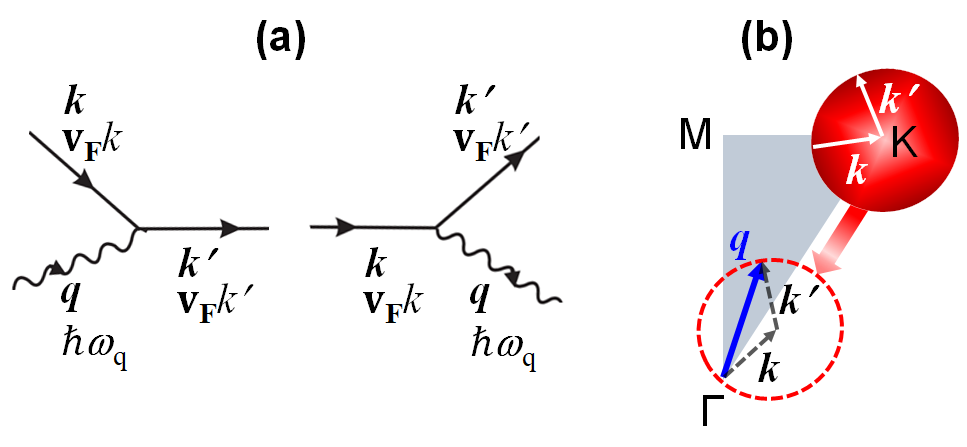}
\caption{The 3-particle e-ph interaction. (a) Unlike metals where only the EP-E* process is considered in 2D Dirac crystals, it is important to also consider the E-E*P* process to derive the crystals' thermal properties. (b) The reciprocal lattice of a honeycomb structured lattice, in which the e-ph interaction occurs. This provides a critical insight into the wave vectors associated with lattice vibrations and their influence on thermal properties.}
\label{Figure1}
\end{figure}

To compute the phonon scattering rate we must first write the interaction Hamiltonian of the e-ph coupling process. The interaction Hamiltonian of a 2D Dirac crystal for the 3-particle e-ph coupling process for in-plane phonons is equal to \cite{jishi2013feynman,taylor2002quantum}:

\begin{equation}
\begin{split}
   H^{3}_{e-ph}=\sum_{q} \sum_{k,k''} \sqrt{\frac{\hbar^{2}}{2m\hbar \omega_{\textbf{q}}}}\cdot q \; \phi_s(q) \; (b^{\dagger}_{q}+b_{q}) \; c^{\dagger}_{k''} c_{k} \; e^{i(\textbf{k}''-\textbf{k}\pm \textbf{q})\cdot \textbf{r}} \; e^{-\frac{i}{\hbar}(E_{k''}-E_{k}\pm \hbar \omega_{\textbf{q}})\cdot t},
\label{The interaction Hamiltonian of the three-particle e-ph process}
\end{split}
\end{equation}

where the $\pm$ sign respectively indicates the 3-particle phonon creation and annihilation processes shown in Fig.~\ref{Figure1}(a). The variable $m$ is the mass of the atom, $\phi_s(q)$ is the screening potential, and $\textbf{k}$ and $\textbf{k}''$ are the wave vectors of the created and annihilated electrons respectively. Furthermore, the fermionic and bosonic creation and annihilation operators are written as $c$, $c^{\dagger}$, and $b$, $b^{\dagger}$ respectively. Further on, the variable $\omega_\textbf{q}$ is the phonons frequency derived using the Komatsu relation \cite{komatsu1951theory,alofi2014theory}, and $E_{k''}$, $E_{k}$ are the energy of the created and annihilated electrons respectively. The energy of acoustic phonons and electrons in 2D Dirac crystals can be written as follow:

\begin{equation}
\begin{split}
  \hbar \omega_\textbf{q}=c q, \quad \quad \quad \quad E_k=\mathrm{v}_{\textbf{F}}k,
\label{phonon, electron energy}
\end{split}
\end{equation}

where $c$ is the phonon velocity and $\mathrm{v}_{\textbf{F}}$ is the Fermi velocity written in energy units. By knowing the Hamiltonian of the e-ph interaction process we can derive the transition rate of the interaction \cite{jishi2013feynman,taylor2002quantum}. The transition rate for the EP-E* process is:

\begin{equation}
\begin{split}
    \gamma^{EP-E^{*}}_{i\to f}=\frac{2\pi}{\hbar} \Big(\frac{\hbar^{2}}{2mc}\Big) \sum_{q}  & q \; \phi^{2}_{s}(q)\; \Big<f_{k}(1-f_{k+q})\Big> \Big<n_{q} \Big> \; \delta(E_{k+q}-E_{k}-cq),
\label{phonon annihilation interaction rate}
\end{split}
\end{equation}

where $n_{q}$ is the Bose-Eienstien statistics of the phonons and $f_{k \pm q}$ is the Fermi-Dirac statistics of the electrons. We further write the transition rate for the E-E*P* process as follow:

\begin{equation}
\begin{split}
    \gamma^{E-E^{*}P^{*}}_{i\to f}=\frac{2\pi}{\hbar} \Big(\frac{\hbar^{2}}{2mc}\Big) \sum_{q}  & q \; \phi^{2}_{s}(q)\; \Big<f_{k+q}(1-f_{k})\Big> \Big<1+n_{q} \Big> \; \delta(E_{k+q}-E_{k}-cq).
\label{phonon creation interaction rate}
\end{split}
\end{equation}

The bounds over the phonon wave number, $q$, in the summation can be calculated using the rules governing the conservation of energy and momentum. By writing the conservation of energy and momentum we find the value of $q$ to be:

\begin{equation}
\left\{
\begin{array}{ll}
		& \textbf{k} + \textbf{q} = \textbf{k}' \\
		& \mathrm{v}_{\textbf{F}} k + cq = \mathrm{v}_{\textbf{F}}k' 
	\end{array}
\right.
\Rightarrow q = \frac{2\mathrm{v}_{\textbf{F}}k(c-\mathrm{v}_{\textbf{F}}\cos{\theta})}{(\mathrm{v}^{2}_{\textbf{F}}-c^{2})}, 
\label{q as a function of v_F and k}
\end{equation}

where $\theta$ is the collision angle between the electron and the phonon. The extreme carrier mobility in 2D Dirac crystals results in the Fermi velocity of the electrons to be much larger than the velocity of the acoustic phonons, $\mathrm{v}_{\mathrm{F}}>>c$. This has been further experimentally confirmed for various 2D Dirac crystals such as graphene \cite{hwang2012fermi}, Weyl semi-metals \cite{lee2015fermi}, silicene \cite{kara2012review}, and borophene \cite{xu2016hydrogenated} where $c < 0.01 \mathrm{v}_{\mathrm{F}}$. Therefore, for $q$ to be positive, $q>0$, the range of the angle $\theta$ should be:

\begin{equation}
\begin{split}
    \cos{\theta} \le 0.01 \Rightarrow \pi \le \theta \le \pi /2 - \varepsilon,
\label{range of theta}
\end{split}
\end{equation}

where $\varepsilon \to 0$. Furthermore, to find the maximum value of the phonon wave number, $q_{max}$, we put $\theta = \pi$ and get the following equation:

\begin{equation}
\begin{split}
    q_{max} = \frac{2\mathrm{v}_{\textbf{F}}k}{(\mathrm{v}_{\textbf{F}}-c)} \approx 2k_\textbf{F} + \varepsilon.
\label{maximum q}
\end{split}
\end{equation}

From Eq.(\ref{maximum q}) we find the range of $q$ in the 2D Dirac crystal to be $0 \le q \le 2k_{\mathrm{F}}+\varepsilon$. This is shown in Fig.~\ref{Figure1}(b).

We further proceed to calculate the e-ph screening potential. Considering the range of $q$ for the 3-particle e-ph interaction process, we note that the conventional method for computing the e-ph screening potential in metals is not applicable to 2D Dirac crystals. This conventional approach typically considers phonons with long wavelengths or short wave numbers, $q>2k_{\mathrm{F}}$, and employs the Thomas-Fermi approximation \cite{ashcroft2022solid} to determine the dielectric response function. However, due to the unique nature of 2D Dirac crystals, this conventional methodology cannot be extended to accurately analyze their e-ph screening potential. The absence of a band gap in 2D Dirac crystals dramatically amplifies the effects of any phonon disturbance in the Fermi surface nesting region, $q \approx 2k_\mathrm{F}$, \cite{yan2020superconductivity,ali2016butterfly} and while the Thomas-Fermi approximation assumes the dielectric response function to have a constant value in the region of $0 \le q \le 2k_{\mathrm{F}}$ this is not the case in 2D Dirac crystals \cite{bahrami2017exchange,calandra2007electron}. To ensure precision in expressing the e-ph screening potential, we incorporate the dynamic dielectric response function. This function is tailored to electrons strongly correlated with acoustic phonons in 2D Dirac crystals, offering a nuanced depiction dependent on the phonon wave number. Particularly, it accommodates phonons with short-dispersive energy wavelengths in the range $0 < q \le 2k_{\mathrm{F}}$. This entails utilizing the dielectric response function derived from the Lindhard model beyond the random phase approximation for e-ph interactions, specifically addressing phonons with short-dispersive energy wavelengths \cite{kazemian2023dynamic}. This meticulous approach ensures a comprehensive representation of the e-ph screening potential in the context of 2D Dirac crystals. Using the appropriate model to calculate the dielectric response function the screening potential can be written as follow:

\begin{equation}
\begin{split}
    \phi_s(\psi)= 
    \frac{1}{\Omega}\Big(\frac{2\pi Q^{2}}{q-2\pi\chi(q)} \Big),
\label{screening potential}
\end{split}
\end{equation}

where $\Omega$ is the unit cell area of the 2D Dirac crystal, $Q$ is the ion charge, and $\chi(q)$ is the dynamic dielectric response function written as a function of the phonon wavenumber, $q$. Knowing the screening potential, we can write the phonon scattering rate, $\tau$, for the EP-E* and E-E*P* process as follow:

\begin{equation}
\begin{split}
    \frac{1}{\tau^{EP-E^{*}}}=\pdv[]{\gamma^{EP-E^{*}}_{i\to f}}{\Big<n_{q} \Big>} = \frac{4\pi^2 \alpha^{2}}{\hbar}  \sum^{2k_{\mathrm F} + \varepsilon}_{q=0} q \; \Big(\frac{1}{q-2\pi\chi(q)} \Big)^{2} \; \Big<f_{k}(1-f_{k+q})\Big> \; \delta(E_{k+q}-E_{k}-cq),
\label{phonon annihilation scattering rate}
\end{split}
\end{equation}

\begin{equation}
\begin{split}
    \frac{1}{\tau^{E-E^{*}P^{*}}}=\pdv[]{\gamma^{E-E^{*}P^{*}}_{i\to f}}{\Big<n_{q+1} \Big>} = \frac{4\pi^2 \alpha^{2}}{\hbar} \sum^{2k_{\mathrm F} + \varepsilon}_{q=0} q \; \Big(\frac{1}{q-2\pi\chi(q)} \Big)^{2} \; \Big<f_{k+q}(1-f_{k})\Big> \; \delta(E_{k+q}-E_{k}+cq).
\label{phonon creation scattering rate}
\end{split}
\end{equation}

Where the variable $\alpha$ is defined as, $\alpha=\frac{1}{\Omega}\big( \frac{\hbar^{2}}{2mc}\big)^{1/2} Q^{2} $, and is dependent on the unit cell area, mass, and the ion charge of the 2D Dirac crystal. The variable $\alpha$ is the variable that we'll use to differentiate between the different 2D Dirac crystals under study. For a 2D Dirac crystal such as graphene with a unit cell area of $\Omega \approx 3.2*10^{-16}$ we have $\alpha \approx 10^{-7}$. By subtracting the phonon scattering rate of the E-E*P* process from the EP-E* process we get the total phonon scattering rate of the 3-particle process and write it as the following:

\begin{equation}
\begin{split}
    \tau^{(3)}=\tau^{EP-E^{*}}-\tau^{E-E^{*}P^{*}}.
\label{total phonon scattering rate}
\end{split}
\end{equation}

In metals with huge Fermi energies, $E_\mathrm{F}$, where $E_\mathrm{F}>>k_B \mathrm{T}$, the Fermi-Dirac distribution of the EP-E* process Eq.(\ref{phonon annihilation scattering rate}) is much larger than the E-E*P* process Eq.(\ref{phonon creation scattering rate}), $\big<f_{k-q}(1-f_{k+q})\big> >> \big<f_{k+q}(1-f_{k-q})\big>$. This leads to the total phonon scattering rate of the 3-particle process to be approximately equal to the phonon scattering rate of the phonon annihilation process, $\tau^{(3)} \approx \tau^{EP-E*}$. However, this is not the case in 2D Dirac crystals in which we have a limited concentration of free electrons and the Fermi energy, $E_\mathrm{F}$, cannot generally be assumed to be much larger than the Boltzmann energy, $k_B \mathrm{T}$. Hence, considering the cancellation of the EP-E* process by the E-E*P* process for low Fermi energies where $E_{\mathrm{F}} \le k_B \mathrm{T}$, it becomes imperative to delve into higher order e-ph interactions. By exploring these higher-order interactions, we can gain a comprehensive understanding of the intricate mechanisms that influence the electronic and thermal properties of 2D Dirac systems. 

\begin{figure}[h]
\centering
\includegraphics[width=.6\columnwidth]{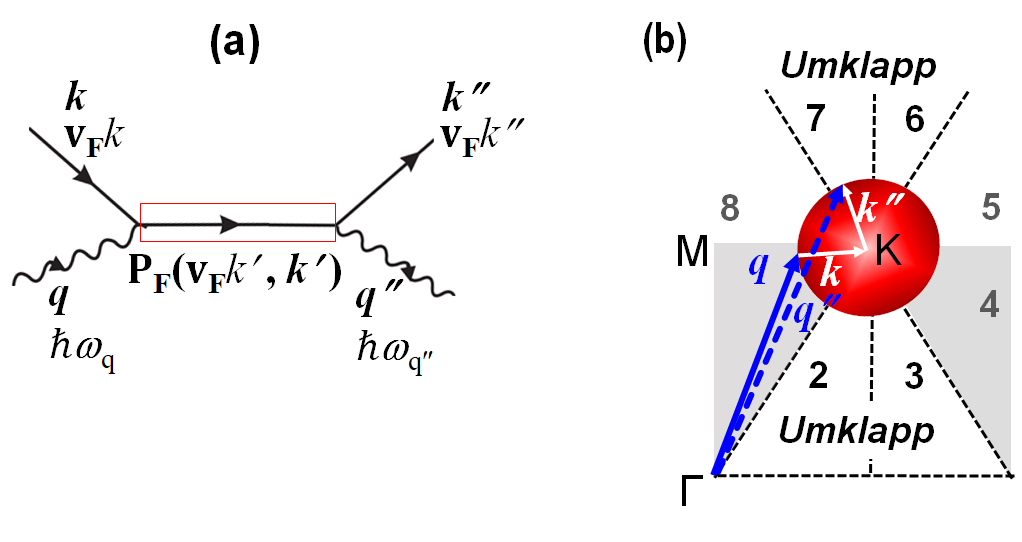}
\caption{(a) The 4-particle e-ph interaction. It consists of the annihilation of a pair of electrons and phonons and the creation of a new pair, EP-E*P*, with a Fermion propagator in the middle. (b) Due to the low $\pi$-$\pi*$ electron energy spacing in the zone-boundary region of 2D Dirac crystal, akin to the energy of acoustic phonons, we assume the 4-particle process occurs at the zone-boundary of the 2D Dirac crystal. The wave vector of the outgoing phonon scattered to the 1st, 4th, 5th, and 8th neighboring lattice sites will remain approximately the same while the wave vector of the phonon scattered to the 2nd, 3rd, 6th, and 7th neighboring lattice sites will be much smaller than the annihilated phonon since its wave vector will be hugely affected by the Umklapp process rendering the process inelastic.}
\label{Figure2}
\end{figure}

The second-order e-ph interaction is a 4-particle process. As shown in Fig.~\ref{Figure2}(a), the 4-particle process consists of the annihilation of a pair of electrons and phonons and the creation of a new pair, EP-E*P*, with a Fermion propagator in the middle. The 4-particle process can also consist of an electron annihilation and phonon creation on the left-hand side of the interaction and an electron creation and phonon annihilation on the right-hand side of the interaction, EP*-E*P, yielding the same result. We will therefore only consider the EP-E*P* process and multiply the final result by two. As shown in Fig.~\ref{Figure2}(b) we assume the e-ph interaction of the 4-particle process occurs at the zone-boundary region, $\mathrm{K}$, of the 2D Dirac crystal. This is because the $\pi$-$\pi*$ electron energy spacing in the zone-boundary region of 2D Dirac crystals is considerably lower, akin to the energy of acoustic phonons, in contrast to the zone-center region \cite{katsnelson2007graphene,geim2010rise}, making the e-ph interaction far more likely to occur at the zone-boundary region. Also, unlike the 3-particle process, in the 4-particle process, the interaction of phonons with zone-boundary electrons does not violate the conservation of momentum. The interaction Hamiltonian of the 4-particle e-ph process shown in Fig.~\ref{Figure2}(a) is written as:

\begin{equation}
\begin{split}
   H^{4}_{e-ph}=\sum_{q,q''} \sum_{k,k',k''} &\sqrt{\frac{\hbar^{2}}{2m\hbar \omega_{\textbf{q}}}}\cdot q \; \phi_s(q) \; b_{q} c^{\dagger}_{k'} c_{k} \; e^{i(\textbf{k}'-\textbf{k}- \textbf{q})\cdot \textbf{r}} \; e^{-\frac{i}{\hbar}(E_{k}'-E_{k}- \hbar \omega_{\textbf{q}})\cdot t} P_{F}(E_{k'},k') \\& \sqrt{\frac{\hbar^{2}}{2m\hbar \omega_{\textbf{q}''}}}\cdot q'' \; \phi_s(q'') \;b^{\dagger}_{q''} c^{\dagger}_{k''} c_{k'} \; e^{-i(\textbf{k}''-\textbf{k}'- \textbf{q}'')\cdot \textbf{r}} \; e^{\frac{i}{\hbar}(E_{k''}-E_{k}- \hbar \omega_{\textbf{q}''})\cdot t},
\label{The interaction Hamiltonian of the four-particle e-ph process}
\end{split}
\end{equation}

where $P_F(E_{k'},k')$ is the Fermion propagator \cite{padmanabhan2018obtaining,taylor2002quantum,jishi2013feynman} and $E_{k'}$ and $k'$ are its energy and momentum. To study the EP-E*P* process we consider the reciprocal lattice of a honeycomb Dirac crystal. As shown in Fig.~\ref{Figure2}(b) we assume the left-hand side of the 4-particle process, e-ph annihilation, to occur in the 1st Brillouin zone while the outgoing electron and phonon can scatter to either the same Brillouin zone or any of its seven neighboring sites. Due to the conservation of momentum, when the outgoing phonon is scattered to any of its neighboring sites, we have to deduct the inverse lattice vector $\Gamma \mathrm{K}$ from the wave vector of the outgoing phonons. This is the so-called Umklapp process. If the outgoing phonon is scattered to the same lattice site, (1st), or any of the 4th, 5th, and 8th neighboring lattice sites its wave vector will remain approximately the same. This is because the created phonon scattered to the 4th, 5th, and 8th neighboring sites will not be affected by the Umklapp process and the e-ph interaction will be approximately elastic. If on the other hand, the outgoing phonon is scattered to the 2nd, 3rd, 6th, and 7th neighboring lattice sites its wave vector will be much smaller than the annihilated phonon since its wave vector will be hugely affected by the Umklapp process and the process will be nearly inelastic. From the interaction Hamiltonian of the 4-particle process, we write the transition rate as follow:

\begin{equation}
\begin{split}
    \gamma^{EP-E^{*}P^{*}}_{i\to f}=&\Big(\frac{2\pi}{\hbar}\Big)^2 \Big(\frac{\hbar^{2}}{2mc}\Big)^{2} \sum_{q,q''} \sum_{k,k',k''} \frac{q \; \big(q''-\Gamma K\big) \; \phi^{2}_{s}(q)\; \phi^{2}_{s}(q''-\Gamma K)}{\Big(k'-k-(c/\mathrm{v}_{\mathrm{F}})q\Big)^{2}}   \\& \Big<f_{k}\big(1-f_{k''}\big)\Big> \Big<n_{q}\big(1+n_{q''}\big) \Big>\; \delta\big(E_{k'}-E_{k}-cq\big)\delta\big(E_{k''}+c(q''-\Gamma K)-E_{k'}\big),
\label{four-particle process interaction rate}
\end{split}
\end{equation}
 
The wave vectors of the phonons scattered to the 1st, 4th, 5th and 8th lattice sites will be elastic, $q''-\Gamma K \to q''$, while the phonons scattered to the 2nd, 3rd, 6th and 7th neighbouring sites will be inelastic, $q''-\Gamma K \to 0$, resulting in $\gamma^{EP-E^{*}P^{*}}_{i\to f} \to 0$. Therefore, half of the processes will destroy the phonon wave vector reducing the phonon scattering rate of the EP-E*P* process to approximately half its total value. We write the phonon scattering rate of the EP-E*P* process as follows: 

\begin{equation}
\begin{split}
    \frac{1}{\tau^{EP-E^{*}P^{*}}}= & \frac{1}{2} \pdv[]{\gamma^{EP-E^{*}P^{*}}_{i\to f}}{\Big<n_{q+1} \Big>} = \\& \frac{1}{2} \Bigg( \frac{16\pi^4 \alpha^{4}}{\hbar} \Bigg) \sum^{q_{max}}_{q_{min}} \;
    \sum^{q''_{max}}_{q''_{min}} \sum^{k_{\mathrm{F}}}_{k,k''=0}   \; \Big<f_{k}(1-f_{k''})\Big> \Big<n_q\Big> \; \delta(E_{k}+cq-E_{k''}-cq''\big) \\& \Big(\frac{1}{q-2\pi\chi(q)}\Big)^{2} \frac{q \; q''}{\Bigg[\Big(k^{2}+q^{2}-2kq\cos{\theta}\Big)^{1/2}-k-(c/\mathrm{v}_{\mathrm{F}})q\Bigg]^{2}}\Big(\frac{1}{q''-2\pi\chi(q'')}\Big)^{2}.
\label{phonon scattering rate of four-particle process}
\end{split}
\end{equation} 

where $\theta$ is the angle of collision between the annihilated phonon and electron and the maximum and minimum of the annihilated and created phonon wave vector can be derived using Fig.~\ref{Figure2}(b). For the minimum phonon wave vector, we have:

\begin{equation}
\begin{split}
   q_{min}=q''_{min}=\Gamma K - k_{\mathrm{F}},
\label{minimum phonon wave vector}
\end{split}
\end{equation} 

and for the maximum phonon wave vector, we have:

\begin{equation}
\begin{split}
   q_{max}=q''_{max}=\sqrt{\big(\Gamma K\big)^{2}+k_{\mathrm{F}}^{2}-2k_{\mathrm{F}}\big(\Gamma K\big)\cos{\pi /3}}.
\label{maximum phonon wave vector}
\end{split}
\end{equation}

We have computed the scattering rate of in-plane phonons, considering both the 3-particle process (refer to Eq. (\ref{total phonon scattering rate})) and the 4-particle process (\ref{phonon scattering rate of four-particle process}). With the obtained phonon scattering rates, our next step involves deriving the thermal conductivity due to e-ph interactions for in-plane phonons in a 2D Dirac crystal. This e-ph thermal conductivity is expressed through the following equation:

\begin{equation}
\begin{split}
    k_{th,(e-ph)}=\frac{1}{2}v^{2}_{gr} \sum_{q} C_{q}\tau,
\label{e-ph thermal conductivity}
\end{split}
\end{equation}

where $v_{gr}$ is the group velocity of acoustic phonons in the 2D Dirac crystal, and $C_{q}$ is the specific heat capacity per unit area. A detailed derivation of $C_{q}$ has been provided in Appendix A. The total thermal conductivity of the 2D Dirac crystal affected by the e-ph interaction can be achieved by employing \textit{Mattiessen's} rule \cite{ziman2001electrons}, a widely used approach in solid-state physics. The expression for the total thermal conductivity can be written as follows:

\begin{equation}
\begin{split}
    \frac{1}{k_{th}} = \frac{1}{k_{th,(0)}} + \frac{1}{k_{th,(e-ph)}},
\label{total thermal conductivity}
\end{split}
\end{equation}

where $k_{th,(0)}$ is the initial thermal conductivity of the undoped Dirac crystal without any e-ph interactions. The employment of numerical methods in different branches of physics and science has become more common in recent years \cite{ezugwu2017contactless,farhani2023bayesian,farhani2022momentum,afrasiabian2023enhanced}. To obtain accurate results, we employ a MATLAB routine to numerically solve the phonon scattering rate and calculate the e-ph thermal conductivity considering both the 3-particle and 4-particle processes. The analytical approach employed in this section enhances our comprehension of first and second-order e-ph interactions within 2D Dirac crystals. While computational techniques like density functional perturbation theory \cite{nomura2015ab}, many-body perturbation theory \cite{marini2015many,faber2012electron}, and Quantum Monte Carlo methods \cite{beyl2018revisiting} often treat e-ph interactions as minor perturbations and may not fully account for strongly correlated interactions, they can also pose challenges when applied to emerging or less-explored materials like recently discovered Dirac crystals. In contrast, the approach detailed in this section is well-suited to address strongly correlated e-ph interactions, adaptable to various 2D Dirac crystal systems, and provides insights into the intricate interplay between different types of 2D Dirac crystals. In the next section, we will present a comprehensive analysis and discussion of the obtained results. By examining the trends and behaviors observed in the calculated e-ph thermal conductivity, we aim to deepen our understanding of the heat conduction mechanisms in the studied 2D Dirac crystal. Through this analysis, we can gain valuable insights into the underlying physics and make informed conclusions regarding the thermal properties of the material.

\section{\label{sec:level3}Result and Discussion}

\begin{figure}[H]
\centering
\includegraphics[width=.6\columnwidth]{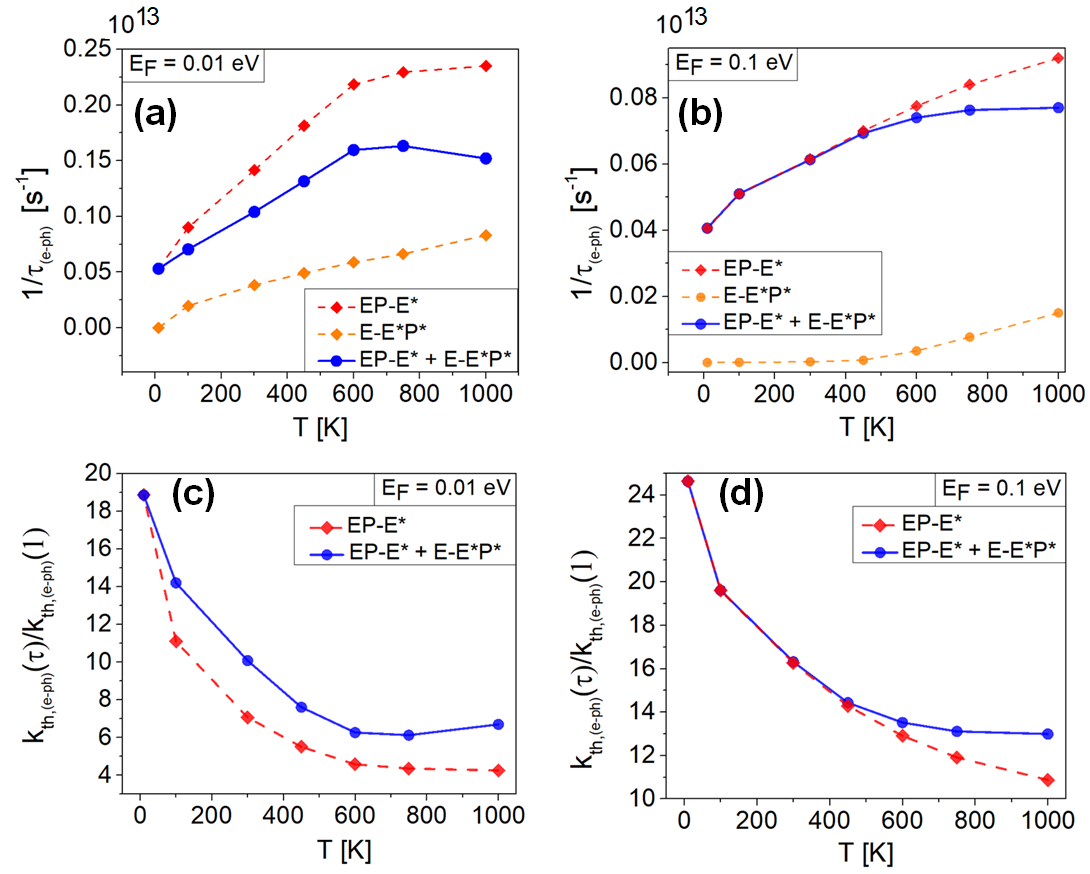}
\caption{(a,b) The inverse phonon scattering rate for in-plane phonons of a 2D Dirac crystal such as graphene with $\alpha \approx 10^{-7}$, as a function of the temperature. Although at low temperatures where $E_\mathrm{F}>>k_B \mathrm{T}$, $\tau^{(3)} \approx \tau^{EP-E*}$ as we increase the temperature the difference between $\tau^{(3)}$ and $\tau^{EP-E*}$ increases. This is because the rate at which the E-E*P* process increases is faster than the EP-E* process, affecting the total transition rate of the 2D Dirac crystal significantly which results in the decrease of the total inverse phonon scattering at high temperatures. (c,d) The e-ph thermal conductivity of a 2D Dirac crystal as a function of the temperature. Similar to the inverse phonon scattering rate the EP-E* is dominant at low temperatures and we have $k^{(3)}_{th,(e-ph)} \approx k_{th,(e-ph)}^{EP-E^{*}}$, however, as we increase the temperature the E-E*P* process becomes important affecting the e-ph thermal conductivity significantly.}
\label{Figure3}
\end{figure}

In order to comprehend the significance of the E-E*P* process in the context of the phonon scattering rate and the e-ph thermal conductivity of 2D Dirac crystals, we initiate our investigation by examining the 3-particle e-ph interactions. In Fig.~\ref{Figure3}(a,b) we have plotted the inverse phonon scattering rate of a 2D Dirac crystal such as graphene with $\alpha \approx 10^{-7}$, as a function of the temperature at two different Fermi energies with.

We observe that at low temperatures where $E_\mathrm{F}>>k_B \mathrm{T}$, $\tau^{(3)} \approx \tau^{EP-E*}$. However, as we increase the temperature the rate at which the E-E*P* process increases is faster than the EP-E* process, affecting the total transition rate significantly. Furthermore, in Fig.~\ref{Figure3}(c,d) we plot the e-ph thermal conductivity of a 2D Dirac crystal for a 3-particle process with $\alpha \approx 10^{-7}$, as a function of the temperature at two different Fermi energies. We observe that although at low temperatures where $E_{\mathrm{F}}>>k_{B}\mathrm{T}$, we have $k^{(3)}_{th,(e-ph)} \approx k_{th,(e-ph)}^{EP-E^{*}}$, as we increase the temperature the E-E*P* process becomes more important affecting the e-ph thermal conductivity significantly. Upon closer examination of Fig.~\ref{Figure3}, our analysis reveals that an intriguing interplay occurs between the E-E*P* process and the EP-E* process in the inverse phonon scattering rate as the temperature rises. Specifically, the former process begins to counteract the latter, ultimately causing the inverse phonon scattering rate to approach zero at elevated temperatures, $E_{\mathrm{F}} << k_{B}\mathrm{T}$. This phenomenon significantly impacts the e-ph thermal conductivity. Consequently, it becomes imperative to investigate higher-order e-ph interactions to gain a comprehensive understanding of the thermal properties of 2D Dirac crystals.

\raggedbottom

\begin{figure}[h]
\centering
\includegraphics[width=.6\columnwidth]{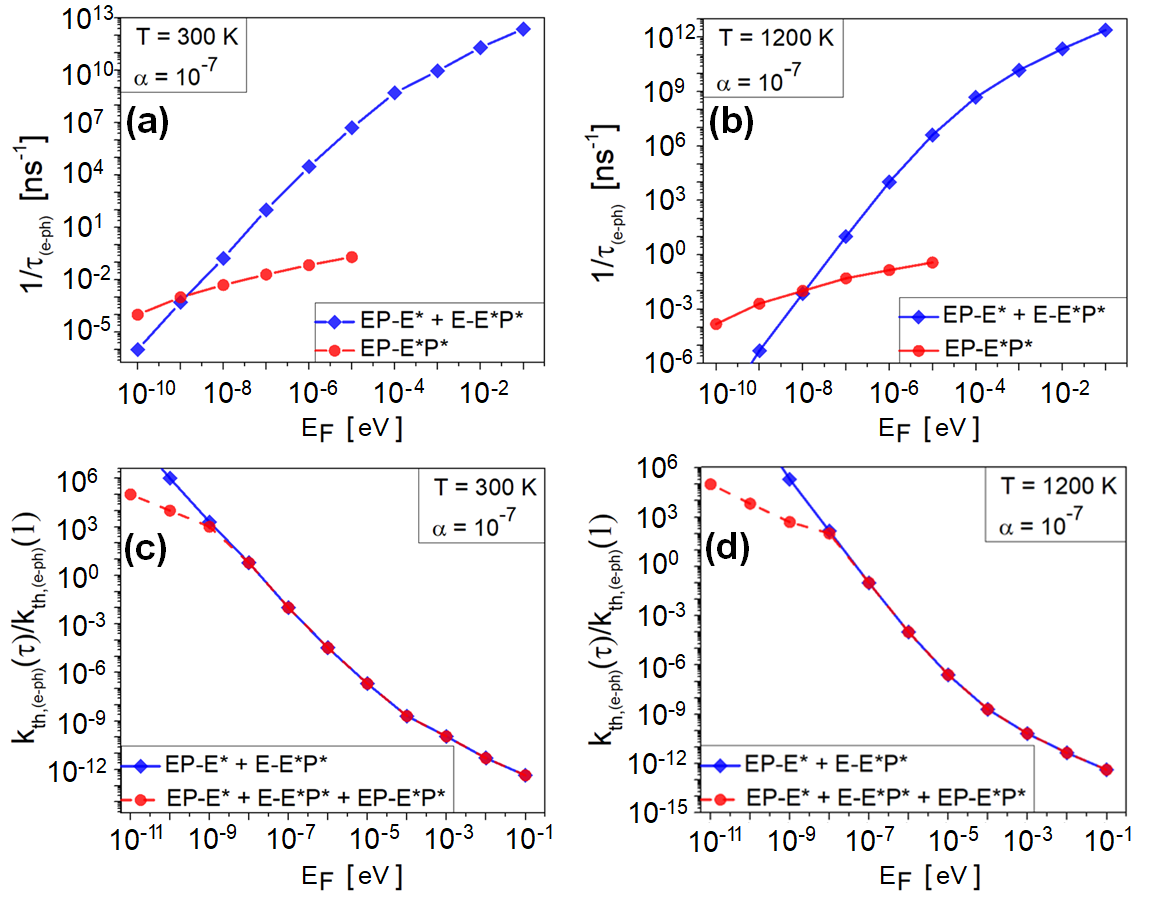}
\caption{(a,b) The relationship between the Fermi energy and the inverse phonon scattering rate for in-plane phonons of a 2D Dirac crystal with $\alpha=10^{-7}$ is observed. When $E_{\mathrm{F}}$ is high, the inverse phonon scattering rate for the 3-particle process is greater than that of the 4-particle process. Conversely, as the Fermi energy decreases, the inverse phonon scattering rate for the 4-particle process becomes more significant. This occurs because, at low $E_{\mathrm{F}}$, the EP-E* process is partially or fully countered by the E-E*P* process, causing the 3-particle process to decrease rapidly compared to the 4-particle process. (c,d) The e-ph thermal conductivity with respect to the Fermi energy of a 2D Dirac crystal with $\alpha=10^{-7}$. We observe an interesting behavior in the e-ph thermal conductivity of the 2D Dirac crystal. When we incorporate the 4-particle process in the calculation of the e-ph thermal conductivity, the rate of increase in slope is significantly reduced at lower Fermi energies compared to when only the 3-particle process is considered. This finding emphasizes the significance of higher-order e-ph interactions in the investigation of 2D Dirac crystals, particularly at low Fermi energies.}
\label{Figure4}
\end{figure}
\raggedbottom

In Fig.~\ref{Figure4}(a,b) we plot the inverse phonon scattering rate as a function of the Fermi energy, $E_{\mathrm{F}}$, for in-plane phonons of a 2D Dirac crystal such as graphene with $\alpha = 10^{-7}$ at temperatures T = 300 K and T = 1200 K. We observe that while the 3-particle process exhibits a higher inverse phonon scattering rate than the 4-particle process at high Fermi energies, a notable shift occurs as we decrease the value of $E_{\mathrm{F}}$. At lower Fermi energies, the inverse phonon scattering rate of the 4-particle process becomes larger and assumes greater significance, warranting careful consideration. The reason for this phenomenon lies in the cancellation of the EP-E* process by the E-E*P* process at low $E_{\mathrm{F}}$ to $k_{B}\mathrm{T}$ ratios. As a result, the 3-particle process experiences a rapid decrease, outpacing the decline of the 4-particle process. We further plot the e-ph thermal conductivity, Fig.~\ref{Figure4}(c,d), as a function of the Fermi energy. Our observations indicate that incorporating the 4-particle process in the analysis of e-ph thermal conductivity results in a notably slower increase in slope at small Fermi wave numbers compared to considering only the 3-particle process. This finding emphasizes the significance of higher-order e-ph interactions in studying 2D Dirac crystals, particularly when examining low Fermi energies. The e-ph thermal conductivity and phonon scattering rate can also be expressed as a function of the carrier concentration instead of the Fermi energy. In 2D Dirac crystals, the relationship between the Fermi energy, $E_{\mathrm{F}}$, and the carrier concentration, $n$, can be described using the following formula:

\begin{equation}
\begin{split}
    E_{\mathrm{F}}= \mathrm{v_{F}} \sqrt{\pi n},
\label{The Fermi energy in terms of the carrier concentration}
\end{split}
\end{equation}

The Fermi energy of 2D Dirac crystals can also be written in terms of energy density. This relationship has been derived in Appendix C of the paper.

Our research is meticulously focused on examining e-ph interactions specifically with in-plane phonons, excluding flexural phonons. Investigating in-plane phonons holds critical importance due to their pronounced role in the e-ph interactions within 2D semiconductors. Compared to flexural phonons, in-plane phonons demonstrate a significantly higher relevance, in the e-ph interactions of these materials highlighting the unique dynamics of their thermal and electrical conductivity \cite{rudenko2019interplay}. Moreover, research indicates that in 2D Dirac crystals, such as graphene, when we have $\sigma_{h}$ symmetry, a distinct lack of coupling exists between electrons and acoustical phonons in the flexural mode \cite{alidoosti2022sigma}. This further underscores the paramount importance of in-plane phonons in understanding the fundamental properties and behaviors of 2D Dirac crystals. Also, the distribution of thermal conductivity across distinct phonon polarization branches in diverse Dirac crystals is contingent upon factors such as sample size and temperature. In the case of graphene nanoribbons, a notable reduction in the significance of flexural phonon modes is observed with rising temperatures—declining from approximately 80\% at T$\approx10$ K to 20\% at 80 K. Beyond T > 100 K, the predominant heat carriers undergo a transition to in-plane phonons, constituting 90\% of the total heat transfer \cite{liu2014anomalous, shen2014size}. This choice aligns with Klemens' theory \cite{klemens2000theory, klemens2001theory}, emphasizing the limited heat-carrying capacity of flexural phonons due to their small group velocities and substantial Gruneisen parameter. Importantly, our consideration of higher-order e-ph interactions is confined to elevated temperatures (T > 300 K), where in-plane phonons overwhelmingly dominate. Comprehensive insights into this aspect are thoroughly discussed in Appendix B, ensuring a nuanced and comprehensive understanding.

\begin{figure}[h]
\centering
\includegraphics[width=.6\columnwidth]{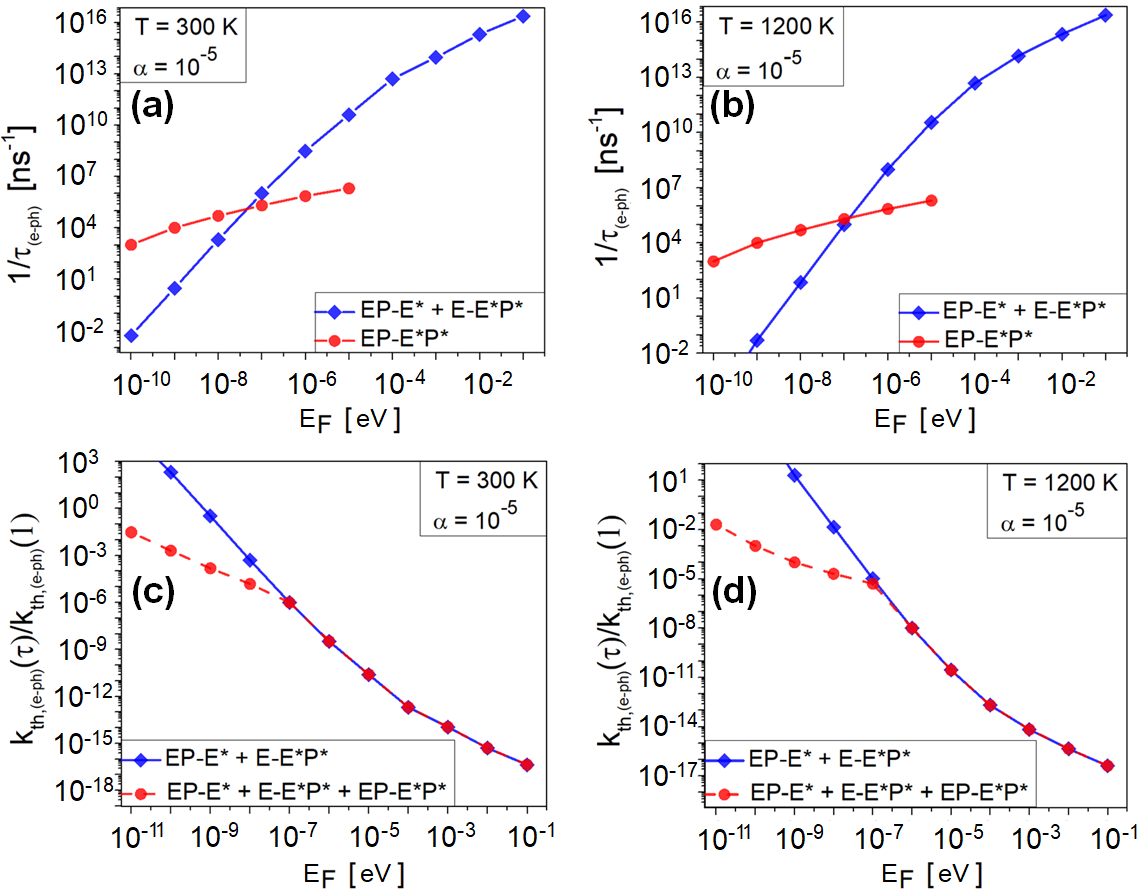}
\caption{(a,b) The inverse phonon scattering rate, as a function of the Fermi energy. At high Fermi energies, the dominant factor influencing the inverse phonon scattering rate is the 3-particle process. However, as we shift to lower Fermi energies, the significance of the 4-particle processes increases, indicating their growing importance in determining the inverse phonon scattering rate. (c,d) The e-ph thermal conductivity, as a function of the Fermi energy. The inclusion of the 4-particle process results in a significant reduction in the rate of increase in slope compared to when only the 3-particle process is taken into account. We notice that increasing the parameter $\alpha$ causes the dominance of the 4-particle process to become more prominent at higher Fermi energies. This phenomenon arises because as $\alpha$ increases, the difference between the strengths of first and second-order e-ph interactions diminishes. Consequently, the probability of 4-particle processes occurring at higher Fermi energies increases as a result of this diminished disparity.}
\label{Figure5}
\end{figure}

Numerous studies have delved into the distinctive thermal characteristics exhibited by 2D Dirac crystals, with graphene as a prominent example \cite{balandin2011thermal,wang2022thermal}. Subsequent investigations have aimed to shed light on the effects of phonon disorder resulting from lattice imperfections, irregularities, as well as variations in crystal size and temperature, on the thermal conductivity of different types of 2D Dirac crystals \cite{liu2014anomalous,nika2017phonons,bae2013ballistic}. In this context, our research delves deeper into unraveling the intricate interplay between e-ph thermal conduction and phonon scattering rate across different 2D Dirac crystals. Here we study the phonon scattering rate and e-ph thermal conductivity for another group of 2D Dirac crystals other than graphene with $\alpha = 10^{-5}$. In Fig.~\ref{Figure5}(a,b) we plot the inverse phonon scattering rate, and in Fig.~\ref{Figure5}(c,d) we plot the e-ph thermal conductivity for in-plane phonons as a function of the Fermi energy. We observe that that similar to 2D Dirac crystals with $\alpha = 10^{-7}$, in 2D Dirac crystals with $\alpha = 10^{-5}$, the inverse phonon scattering rate and $k_{th,(e-ph)}$ are dominated by 3-particle processes at high Fermi energies while at low Fermi energies, the 4-particle processes become more important. By further comparing Fig.~\ref{Figure4} and Fig.~\ref{Figure5}, it is evident that raising the value of $\alpha$ leads to the prevalence of the 4-particle process at higher Fermi energies. This phenomenon occurs due to the diminishing disparity between first and second-order e-ph interactions as $\alpha$ increases, thereby increasing the likelihood of 4-particle processes at higher Fermi energies.  

\section{\label{sec:level4}Conclusion}

In this paper, we present a theoretical framework for computing the phonon scattering rate and the e-ph thermal conductivity in 2D Dirac crystals for in-plane phonons. To this end, we derive an accurate expression of the e-ph interactions for acoustical phonons with short-dispersive energy wavelengths necessary for studying the thermal properties of 2D Dirac crystals. We then derive the phonon scattering rate and the e-ph thermal conductivity for the 3-particle process. Additionally, our research explores the impact of Umklapp scattering on the e-ph thermal conductivity of 2D Dirac crystals at elevated temperatures and the complex interplay between e-ph thermal conduction and phonon lifetimes across different 2D Dirac crystals. Our findings demonstrate a distinction between metals and 2D Dirac crystals in terms of the considered processes. While metals focus solely on the e-ph decay and the creation of a new electron, EP-E*, in the 3-particle process, 2D Dirac crystals necessitate the inclusion of processes involving electron decay and the creation of a new electron and phonon, E-E*P*. Our study reveals that at elevated temperatures and low Fermi energies, the E-E*P* process has the potential to partially or entirely nullify the EP-E* process. Consequently, it becomes imperative to investigate higher-order e-ph interactions in order to comprehend the overall dynamics accurately. Ultimately, our findings demonstrate the impact of the 4-particle process, EP-E*P*, on both phonon scattering rate and e-ph thermal conductivity. Specifically, we observe its heightened prominence compared to the 3-particle process at low Fermi energies and high temperatures. Future research in the field could explore two intriguing aspects: firstly, the potential engineering of thermal rectifiers in 2D Dirac crystals through tailored e-ph interactions, enabling preferential heat flow in one direction. Secondly, investigating how the electronic band structure of 2D Dirac crystals, characterized by elements such as Dirac cones and electronic dispersion, impacts e-ph interactions and thermal transport represents another promising avenue for exploration.

\centerline{***}
\paragraph*{\bf Acknowledgments}
We would like to thank Dr. Ghazal Farhani for helping us with the numerical calculations and for her helpful discussions in analyzing the results. We would further like to thank Dr. Sadra Hajkarim for his helpful discussions. Finally, we would like to thank Dr. James Gaudet for lending us his valuable resources for writing this paper. This work was supported by the Natural Science and Engineering Research Council of Canada (NSERC) through a Discovery Grant (RGPIN-2020-06669). We acknowledge the Anishinaabek, Haudenosaunee, L¯unaap´eewak and Attawandaron peoples, on whose traditional lands Western University is located.%

\appendix

\section{Appendix: Specific heat capacity}

The specific heat of the crystal lattice can be written as \cite{ashcroft2022solid}:

\begin{equation}
\begin{split}
   C_{q}=\pdv[]{}{T}\int \frac{d\textbf{q}}{4\pi^2} \frac{\hbar \omega_{\mathrm{q}}}{e^{\hbar \omega_{\mathrm{q}}/k_B T}-1} \, .
\label{specific heat}
\end{split}
\end{equation}

The Debye model replaces all branches of the vibrational spectrum with linear dispersion relations as follows:

\begin{equation}
\begin{split}
   \hbar \omega_{\mathrm{q}}=cq.
\label{phonon energy}
\end{split}
\end{equation}

Additionally, the integration in Eq. (\ref{specific heat}) across the first Brillouin zone is substituted with an integration over a sphere with a radius of $q_D$. This radius is selected to encompass precisely $N$ permissible wave vectors, where $N$ corresponds to the number of ions present in the crystal. $q_D$ is the measure of the inverse interparticle spacing and its relation with the unit surface area can be written in the following manner:

\begin{equation}
\begin{split}
   n=\frac{N}{\Omega}=\frac{q_{D}^{2}}{4 \pi},
\label{total number of ions per unit surface}
\end{split}
\end{equation}

where $n$ is the total number of ions per unit surface area of the crystal. We write $q_D$ as follow:

\begin{equation}
\begin{split}
   q_{D}=2\sqrt{\frac{\pi N}{\Omega}}.
\label{calculating q_D}
\end{split}
\end{equation}

We can therefore write the specific heat capacity as follow:

\begin{equation}
\begin{split}
   C_{q}=\pdv[]{}{T}\int_{0}^{2\sqrt{\frac{\pi N}{\Omega}}} \frac{2 \pi dq}{4 \pi^{2}} \frac{cq}{e^{cq/k_B T}-1} \, .
\label{specific heat final}
\end{split}
\end{equation}

The integral in Eq. (\ref{specific heat final}) can be further solved numerically and by applying different approximations for low, intermediate and high temperatures \cite{ashcroft2022solid}. 

\section{Appendix: Thermal conductivity along different phonon branches of graphene nanoribbon as a function of temperature}

Graphite crystal is conceptualized as a system comprising thin elastic plates, each separated by a constant distance. The carbon atoms within adjacent atomic layers are spaced at a distance of 3.40 Å. In a hexagonal crystal, there exist five independent elastic constants, namely $c_{11}$, $c_{12}$, $c_{13}$, $c_{33}$, and $c_{44}$. When addressing vibrations of layers with atomic displacements confined to their plane, only two independent constants, $c_{11}$ and $c_{12}$, need consideration. Consequently, a hexagonal layer can be viewed as a continuous isotropic medium. The velocities of longitudinal and transverse waves within this medium are as follows:

\begin{equation}
\begin{split}
   v_{l}=\sqrt{\frac{E}{2 \rho (1+\sigma)}},
\label{longitudinal speed}
\end{split}
\end{equation}

\begin{equation}
\begin{split}
   v_{t}=\sqrt{\frac{E}{\rho (1-\sigma^{2})}},
\label{transversal speed}
\end{split}
\end{equation}

where $\rho$, $E$, and $\sigma$ are the volume density, the Young’s modulus, and the Poisson’s ratio respectively. Utilizing the semi-continuum model outlined in \cite{komatsu1951theory}, Nishira and Ivata \cite{nihira2003temperature} obtained analytical expressions for the different phonon frequency branches we have:

\begin{equation}
\begin{split}
   \omega^{2}_{LA}=v^{2}_{l}\Big(q^{2}_{x}+q^{2}_{y}\Big) + \frac{4\zeta}{c^{2}}\sin^{2}{\Big(\frac{cq_{z}}{2}\Big)},
\label{Longitudinal frequency}
\end{split}
\end{equation}

\begin{equation}
\begin{split}               
    \omega^{2}_{TA}=v^{2}_{t}\Big(q^{2}_{x}+q^{2}_{y}\Big) + \frac{4\zeta}{c^{2}}\sin^{2}{\Big(\frac{cq_{z}}{2}\Big)},
\label{Transversal frequency}
\end{split}
\end{equation}

\begin{equation}
\begin{split}               
    \omega^{2}_{ZA}=b^{2}\Big(q^{2}_{x}+q^{2}_{y}\Big)^{2} + 4\mu^{2}\sin^{2}{\Big(\frac{cq_{z}}{2}\Big)} + \zeta\Big(q^{2}_{x}+q^{2}_{y}\Big),
\label{Flexural frequency}
\end{split}
\end{equation}

where c is the interlayer spacing, b is the
bending elastic parameter, $\zeta = c_{44}/\rho$ and $\mu = \sqrt{c_{33}/(\rho c^{2})}$. Understanding the distinct phonon frequency branches and the phonon scattering rate \cite{shen2014size} allows us to compute the thermal conductivity for each branch. While analytical integration for thermal conductivity in graphene is not feasible, it can be effectively accomplished through the Monte Carlo sampling method \cite{xie2013phonon}, expressed as:

\begin{equation}
    k_{\lambda} =
\left\{
\begin{array}{ll}
		\frac{k_{B}\omega_{\lambda}}{2 \pi \delta} \frac{1}{N}  \sum_{i=1}^{N} \frac{\big(\hbar \omega/k_{B} T\big)^{2} e^{\hbar \omega/k_{B} T}} {\big(e^{\hbar\omega/k_{B}T}-1\big)^{2}} \omega_{\lambda} \cos^{2}\theta \tau_{\lambda}  & \lambda = LA,TA\\
		\frac{k_{B}\omega_{\lambda}}{ \pi \delta} \frac{1}{N}  \sum_{i=1}^{N} \frac{\big(\hbar \omega/k_{B} T\big)^{2} e^{\hbar \omega/k_{B} T}} {\big(e^{\hbar\omega/k_{B}T}-1\big)^{2}} \omega_{\lambda} \cos^{2}\theta \tau_{\lambda}  & \lambda = ZA
	\end{array}
\right.
\label{Difference between the two Fermi-Dirac distributions}
\end{equation}

Here, N represents the sampling number, and it is crucial for its value to be sufficiently large, such as $10^6$, to enhance accuracy and minimize variance. The variable $\tau_{\lambda}$ is the phonon scattering rate. Matthiessen’s rule, premised on the independence of various scattering mechanisms, guides the consolidation of diverse phonon interactions. This consolidation encompasses the three-particle phonon-phonon interaction, which, the 3-particle phonon-phonon scattering rate, $\tau_{3-ph}$,  as per time-dependent perturbation theory \cite{roufosse1973thermal}, is expressed as follows:

\begin{equation}
   \tau_{3-ph}=\frac{M v^{2}_{\lambda} \omega_{D,\lambda}}{\gamma^{2}_{\lambda} k_{B} T \omega^{2}} e^{\Theta_{\lambda}/3T},
\label{3-particle phonon-phonon scattering}
\end{equation}

where $M$ is the mass of a graphene unit cell, $\gamma_{\lambda}$ is the Gruneissen parameter which controls the strength of the phonon-phonon scattering process for each branch, $\omega_{D,\lambda}$ is the Debye frequency, and $\Theta$ is the Debye temperature for each branch. We graph the thermal conductivity of a graphene ribbon with specific dimensions across various phonon branches as a function of the temperature. A notable trend emerges, revealing a substantial decrease in the significance of flexural phonon modes as temperatures rise—from $80\%$ at T$\approx10$ K to $20\%$ at 80 K. Beyond T > 100 K, the primary heat carriers shift to in-plane phonons, constituting $90\%$ of the total heat transfer. This finding aligns with outcomes reported in other studies \cite{liu2014anomalous, shen2014size}. Consequently, in the realm of elevated temperatures (T > 300 K), characterized by higher-order e-ph interactions, in-plane phonons overwhelmingly dominate the heat transfer process.

\begin{figure}[h]
\centering
\includegraphics[width=.6\columnwidth]{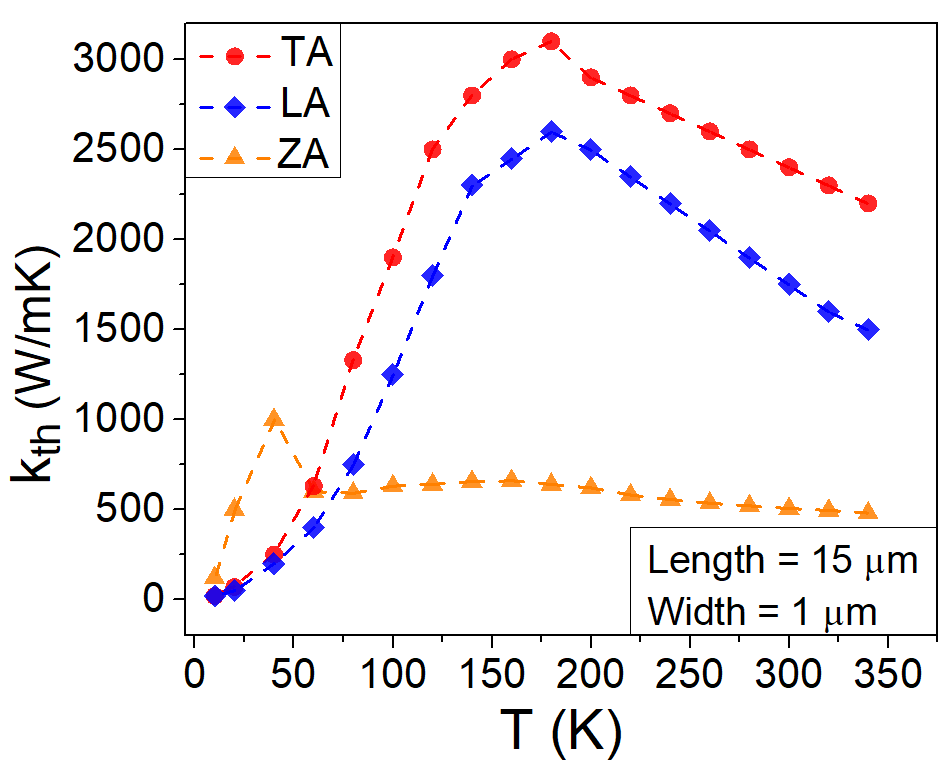}
\caption{The temperature-dependent thermal conductivity profiles of distinct phonon branches in a graphene ribbon. Noticeably, with an elevation in temperature beyond T > 100 K, the thermal conductivity becomes increasingly reliant on the in-plane branches of the graphene ribbon.}
\label{Figure6}
\end{figure}

\section{Appendix: Energy Density of 2D Dirac crystals}

The energy density of 2D Dirac crystals depends on various factors, including the crystals' electronic band structure, Fermi energy, and temperature. In a 2D Dirac material, such as graphene, the electronic band structure near the Fermi level can be described by Dirac cones, where the dispersion relation follows the form:

\begin{equation}
   E_{\mathrm{k}}= \mathrm{v_{F}} k,
\label{Fermi energy near the Dirac cone}
\end{equation}

To calculate the energy density, $g(E)$, of such a crystal, one would integrate the density of states over the energy range of interest. The energy density near the Fermi energy can be written as:

\begin{equation}
   g(E)= 2 \pi k \Big(\frac{dE}{dk}\Big)^{-1}=\frac{ 2 \pi E_{\mathrm{F}}}{\mathrm{v_F}^2}.
\label{Energy Density}
\end{equation}

It is essential to note that some calculations may necessitate more intricate considerations, including the impact of impurities, temperature fluctuations, and interactions with external fields. Moreover, the energy density is subject to variations contingent upon the nuanced properties inherent to the particular Dirac crystal under examination.

\bibliographystyle{unsrt}
%\bibliography{citation}
{\footnotesize
\bibliography{PINN.bib}}

%. This is because when $E_\mathrm{F}>>k_B \mathrm{T}$ the Fermi-Dirac distribution function, $f_{k\pm\frac{q}{2}}$, of an EP-E* process is much larger than the E-E*P* process, $\big<f_{k+\frac{q}{2}} (1-f_{k-\frac{q}{2}})\big> >> \big<f_{k-\frac{q}{2}} (1-f_{k+\frac{q}{2}})\big>$

%Furthermore, by applying the conservation of momentum and the conservation of energy in 2D Dirac crystals for the e-ph three-particle interaction process we show that the range of the phonon wavenumber is between $0\le q \le 2k_\mathrm{F} + \varepsilon$,  where $\varepsilon \to 0$.

\end{document}